\documentclass[traditabstract]{aa}

\usepackage{graphicx}
\usepackage[]{subfigure}
\usepackage{epsfig}
\usepackage{graphicx}
\usepackage{enumerate}
\usepackage{txfonts}
\usepackage{natbib}
\bibpunct{(}{)}{;}{a}{}{,} % to follow the A&A style

\newcommand{\nc}{\newcommand}
\nc{\mb}{\mathbf}
\nc{\LCDM}{$\Lambda$CDM}
\nc{\wcdm}{$w$CDM+$k$}
\nc{\mbf}{\mathbf}
\begin{document}

\title{Reactor sterile neutrinos, dark energy and the age of the universe}
\author{Jostein R. Kristiansen
\and \O ystein Elgar\o y}
\institute{Institute of theoretical astrophysics, University of Oslo, Box 1029, 0315 Oslo, NORWAY}
\date{\today}

\abstract{There are indications that the neutrino oscillation data from reactor experiments and the LSND and MiniBooNE experiments show a preference for two sterile neutrino species, both with masses in the eV region.  We show that this result has a significant impact on some important cosmological parameters.  Specifically, we use a combination of CMB, LSS and SN1A data and show that the existence of two light, sterile neutrinos would rule out the cosmological constant as dark energy at 95\% confidence level, and lower the expansion age of the universe to $12.58\pm 0.26$ Gyr.}


\titlerunning{Sterile neutrinos, dark energy and the age of the universe} 
\authorrunning{J. R. Kristiansen \& \O. Elgar\o y}

\maketitle

\section{Introduction}   

It is no news that the properties of neutrinos impact the history of the universe.  For example, the first limits on the number of neutrino flavours came from Big Bang Nucleosynthesis (BBN) \citep{steigman:1977}.  And as measurements of the statistical distribution of matter in the universe and the anisotropies in the cosmic microwave background (CMB) have improved, it has become possible to put increasingly 
stringent upper bounds on the sum of the neutrino masses \citep{komatsu:2011,thomas:2010}.  The strongest bounds result, of course, when one starts from the simplest cosmological model with a handful of parameters fitted to a selection of the most important data sets and then includes the sum of the neutrino masses as an additional degree of freedom.  

However, there are good reasons to take the cosmological mass limits with a grain of salt.  For one thing, the cause of the apparent accelerated expansion of the universe is unknown.  Although the cosmological constant is consistent with all existing data \citep{komatsu:2011}, we cannot exclude alternative explanations like scalar fields \citep{mota} , modified gravity \citep{tsujikawa:2010} or void models \citep{mattsson:2010}.  Furthermore, neutrino experiments suggest that the neutrino sector of the Standard Model of particle physics may be more complicated and interesting than the simplest picture with three massive flavour states consisting of a superposition of three mass eigenstates \citep{aguilar:2001,aguilar-arevalo:2010,mueller:2011} .   In this paper our point of departure is the fact that the simplest realizations of massive neutrinos do not seem to explain the results of experiments with neutrinos from nuclear reactors and the LSND and MiniBooNE experiments.  Recently it was suggested that two sterile neutrinos, neutrinos that only interact gravitationally with matter, give a good description of the data if their masses are in the eV-range \citep{kopp:2011}.

The ability of present and future 
cosmological data sets to constrain light sterile neutrinos was investigated in { \citep{gonzalez:2010}}, \citet{calabrese:2011}, and \citet{giusarma:2011}.  We choose to approach the problem from the opposite direction: given the uncertain nature of the dark energy, and the somewhat model-dependent interpretation of cosmological data, we would argue, like we have done in the past \citep{kristiansen:2008,kristiansen:2010}, that neutrino experiments have greater authority than cosmology.  If the latter find that light, sterile neutrinos are required, cosmologist have to find room for them in their models.  Knowing full well that the question is far from settled, we nevertheless find it worthwhile to consider what changes in the cosmological concordance model a scenario with 3 active and 2 light, sterile neutrinos lead to.  We will therefore investigate the scenario where the proposal in \citep{kopp:2011} is assumed to be correct and factor their result into an analysis of current cosmological data.  

Our paper is organized as follows.  In section II and III we summarize the theoretical background and describe our method, but briefly since it follows procedures that are standard in the literature.  Section IV is the most important section where we present our results and the inferences we draw from them.  We summarize and conclude in section V.  

\section{Sterile neutrinos and cosmology}

The number of neutrino species with masses below the GeV scale and
that couple to the Z$^{0}$ boson, i.e., interact weakly, was
determined to be $2.984\pm 008$ from LEP data \citep{nakamura:2010}.  
If there are more
neutrino types than the three we already know about, they must be very
heavy, or couple to gravity only, or both.  Neutrinos that do not
participate in the weak interaction are known as sterile.  They
appear in the so-called seesaw mechanism \citep{zuber:2004} for generating small
neutrino masses, and are there typically very heavy, much heavier than
the electroweak scale, in order to explain the smallness of the masses
of the ordinary, active neutrinos.  

However, as long as it only interacts gravitationally there are no a
priori constraints on the mass of a putative sterile neutrino.
Sterile neutrinos with keV masses have been of great interest as dark
matter candidates in cosmology \citep{kusenko:2009} .  And recently it has been suggested
that one or two sterilie neutrinos with masses of a few eV lie behind
some puzzling features in neutrino oscillation experiments
\citep{kopp:2011}.  Ever since the LSND experiment found indications of
$\overline{\nu}_{\mu}$-$\overline{\nu}_{\rm e}$ transitions \citep{lsnd:1998}, there
have been speculations about the existence of a light, sterile
neutrino.  The MiniBooNE experiment \citep{aguilar-arevalo:2010} 
provided support for the LSND
result, but found no evidence for oscillations in the
$\nu_{\mu}$-$\nu_{\rm e}$ channel.  A recent re-evaluation\citep{mueller:2011} of the
expected antineutrino flux from nuclear reactors hint at neutrino
oscillations over length scales of tens to hundreds of meters.  All of
these results have been shown to be accommodated within a model with
two eV-mass sterile neutrinos that are quite strongly mixed with
electron-type neutrinos.  A single sterile species is compatible with
all the results except the negative MiniBooNE result for the
$\nu_{\mu}$-$\nu_{\rm e}$ channel.  We will take the best-fit models
of \citep{kopp:2011} as our point of departure, investigating both the models with one and two sterile neutrino species.     

An important question when we turn to the cosmological implications of
these two scenarios is whether these light, sterile neutrinos were
thermalized in the early universe.  We will assume that they were,
since several studies \citep{hamann:2010,melchiorri:2009,kainulainen:1990}  suggest that this was the case for the masses and mixing parameters we consider.  This means that standard relation
between the sum of the neutrino masses and their contribution to the
cosmic mass density parameter applies. 

Adding two light sterile neutrinos  may cause some problems with BBN  
since the increased relativistic energy density results in a larger 
neutron-to-proton ratio, and leads to an increased He$/$H  mass 
fraction.
 For
example, a recent analysis \citep{mangano:2011} found that BBN
constrains the number of relativistic degrees of freedom to be $N_{\rm
eff} < 4.2$ at 95 \% confidence.  So the 3+1 model is just within the
bounds, and the 3+2 model is just outside it.  If the neutrino oscillation data ends up pointing unequivocally to the existence of two light, sterile neutrinos, this would mean that the standard BBN scenario has to be modified.  This is beyond the scope of this paper.      

In contrast to the relatively tight upper bound on $N_{\rm eff}$ from BBN, several recent papers have suggested that additional relativistic species are allowed, and in fact preferred, by a wide range of cosmological data.  Specifically, $N_{\rm eff} = 5$ is within the allowed region \citep{hamann:2010, dunkley:2010, giusarma:2011}. The allowed mass range of the sterile neutrinos has also been studied. In \citet{hamann:2010} a scenario with massless flavour neutrinos and two thermalized sterile neutrinos with a common mass was investigated. Using a combination of different cosmological probes, including the CMB and the galaxy power spectrum, they found a    95 \% upper bound of 0.45 eV  for each sterile neutrino. For a single sterile neutrino the upper bound was  0.48 eV.  There is some tension between these limits and the best-fit masses of the sterile neutrinos that we adopt from \citet{kopp:2011}. However, we note that the limits in \citet{hamann:2010} are derived assuming a flat universe with a cosmological constant, while we will allow for both spatial curvature and a dark energy equation of state, $w \neq -1$ in our analysis. 

\section{Method}

We study two different cosmological scenarios, the standard flat \LCDM{} model, and an extended model where we allow the dark energy equation of state and the spatial curvature of the universe to vary, which we will refer to as \wcdm. For both of these models we estimate the model parameters when including zero, one (1 $\nu_s$ ) or two (2 $\nu_s$ ) sterile neutrino species. 
We assume the sterile neutrinos to be fully thermalized, and adopt the best-fit masses for the sterile neutrinos from reactor experiments found in \cite{kopp:2011}, that is $m_{\nu_s} = 1.33$eV in the 1 $\nu_s$ scenario, and $m^{(1)}_{\nu_s} = 0.68$eV and  $m^{(2)}_{\nu_s} = 0.94$eV in the 2 $\nu_s$ scenario. We assume the three species of flavour neutrinos to be massless, which should be a good approximation when the sterile neutrino masses are in the high end of their cosmologically allowed mass range (see e.g. \citet{giusarma:2011}). 

We use a modified version of the publicly available cosmological Markov chain Monte Carlo sampler CosmoMC \citep{lewis:2002} to compute the parameter limits. For the \LCDM{} model we vary the parameter set \{$\omega_b$, $\omega_c$, $\theta$, $\tau$, $n_s$, $\ln 10^{10} A_S$\}, and for the \wcdm{} model we also include $w$ and $\Omega_k$ as free parameters. $\omega_b$ and $\omega_c$ are the physical baryon and cold dark matter densities, respectively. $\theta$ is the ratio of the sound horizon to the angular diameter distance, $\tau$ is the optical depth, $n_s$ and $A_s$ are the primordial scalar spectral index and amplitude (at $k = 0.05 \textrm{Mpc}^{-1}$). $w$ denotes the dark energy equation of state (assumed to be constant), and $\Omega_k$ is the curvature density. For exact parameter definitions we refer to the CosmoMC code.  We marginalize over the SZ amplitude. All the listed parameters are given flat priors. 

We also use two different combinations of data sets. First, we only use CMB data from the WMAP 7 year data release \citep{komatsu:2011, larson:2011}, which we will refer to as WMAP7. Then we also include data on large scale structures from the Sloan Digital Sky Survey DR7 luminous red galaxy sample \citep{reid:2009}, Supernova 1A data from SDSS-II \citep{kessler:2009} and a prior on the Hubble parameter of $H_0 = 73.8 \pm 2.4 \textrm{km s}^{-1} \textrm{Mpc}^{-1}$ \citep{riess:2011}. We will refer to this combination of data sets as WMAP7++. For the \wcdm{} model we only use the WMAP7++ data sets, as WMAP7 data alone have very little constraining power for this extended parameter space.  

\section{Results}

\begin{figure*}[htb]
\center
\includegraphics[width=1.4\columnwidth]{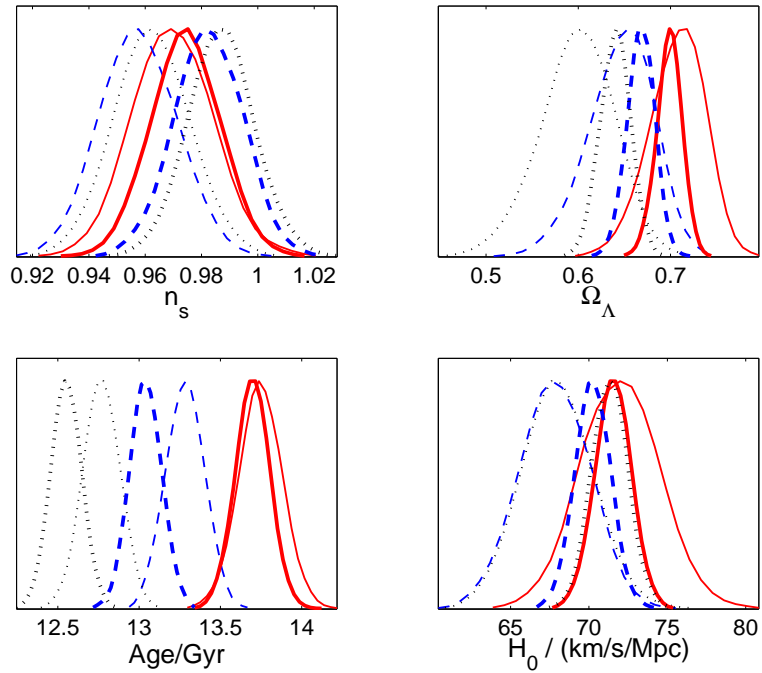}
\caption{Marginalized parameter distributions for \LCDM{} with and without additional sterile neutrinos. Solid, red lines: No sterile neutrinos. Dashed, blue lines: 1 sterile neutrino. Dotted, black lines: 2 sterile neutrinos. Thin lines show results from using WMAP7 data only. Results from WMAP7++ are shown with thick lines. }
\label{fig:1D_LCDM}
\end{figure*}

In Figure \ref{fig:1D_LCDM} we show 1D marginalized probability distributions for a few selected parameters for the \LCDM, \LCDM+1$\nu_s$, and \LCDM+2$\nu_s$ cases, and the corresponding numerical limits are given in Table \ref{tab:lcdm}. 

Interestingly, we see that including sterile neutrinos with their masses fitted to reactor experiments will shift the age of the universe significantly. While \LCDM{} favours an age of the universe of  $13.75 \pm 0.13$ Gyr with WMAP7, the inclusion of 2 sterile neutrinos leads to a preferred age of only $12.77 \pm 0.11$ Gyr. For the WMAP7++ data, the corresponding age estimates are $13.70 \pm {0.10}$Gyr and $12.55 \pm 0.09$ Gyr. The reason for these large shifts can be found in the corresponding shifts in the dark energy density, $\Omega_\Lambda$. The shift in $\Omega_\Lambda$ can be understood by considering changes in the time of matter-radiation equality, $t_{\textrm{eq}}$. At the time of equality the sterile neutrinos were still relativistic, thus they contributed to the relativistic (radiation) energy denisity, and keeping all other parameters constant, additional sterile neutrinos will shift $t_\textrm{eq}$ to later times. The CMB power spectrum is quite sensitive to $t_\textrm{eq}$ (see e.g. \citet{lesgourgues:2006}), and to shift $t_\textrm{eq}$ back, the matter density must be increased. When we require a flat universe, this will lead to a reduction of $\Omega_\Lambda$ and thus a younger universe.

%\begin{widetext}
%\end{widetext}

\begin{table*}[htb]
\begin{center}
\begin{tabular}{crlrlrl}
\hline \hline
 Parameter& \multicolumn{2}{c}{\LCDM} &\multicolumn{2}{c}{\LCDM+1$\nu_s$}& \multicolumn{2}{c}{\LCDM+2$\nu_s$} \\
\hline
\hline
\multicolumn{7}{c}{WMAP7} \\
\hline
$n_s$ & \quad0.970 & $\pm$0.014 & \quad 0.958 & $\pm$ 0.013 & \quad0.962 & $\pm$ 0.013 \\
$\Omega_\Lambda$ & 0.707 & $\pm$0.030 & 0.646 & $\pm$ 0.035 & 0.598 & $\pm$ 0.039 \\
Age (Gyr) &13.75&$\pm$0.13& 13.28&$\pm$0.12& 12.77&$\pm$0.11 \\
$H_0$ (km/s/Mpc) &72.0&$\pm$2.5&67.8&$\pm$2.2& 68.0& $\pm$ 2.2\\
${ \Delta \chi^2}$ &  0 &&  -3.4  &&  -3.6& \\ 
\hline
\multicolumn{7}{c}{WMAP7++}\\
\hline
$n_s$ & \quad0.974 & $\pm$0.012 & \quad 0.982 & $\pm$ 0.012 & \quad0.987 & $\pm$ 0.012 \\
$\Omega_{DE}$ & 0.699 & $\pm$0.014 & 0.668 & $\pm$ 0.015 & 0.640 & $\pm$ 0.016 \\
Age (Gyr) &13.70&$\pm$0.10& 13.04&$\pm$0.10& 12.55&$\pm$0.09 \\
$H_0$ (km/s/Mpc) &71.5&$\pm1.1$&70.2&$\pm$1.1&71.3&$\pm$1.1\\
${ \Delta \chi^2}$ &  0 &&  -24.7  && -22.6& \\ 
\hline
\end{tabular}
\caption{One-dimensional marginalized parameter limits for \LCDM{} with and without sterile neutrinos. The errors shown are the 1$\sigma$ deviations from the mean value.  We also show the $\Delta \chi^2$ values for the different models where ${  \Delta \chi^2_{\rm model} = \chi^2_{\Lambda {\rm CDM}} - \chi^2_{\rm model}}$.}
\label{tab:lcdm}
\end{center}
\end{table*}

\begin{figure*}[htb]
\center
\includegraphics[width=1.4\columnwidth]{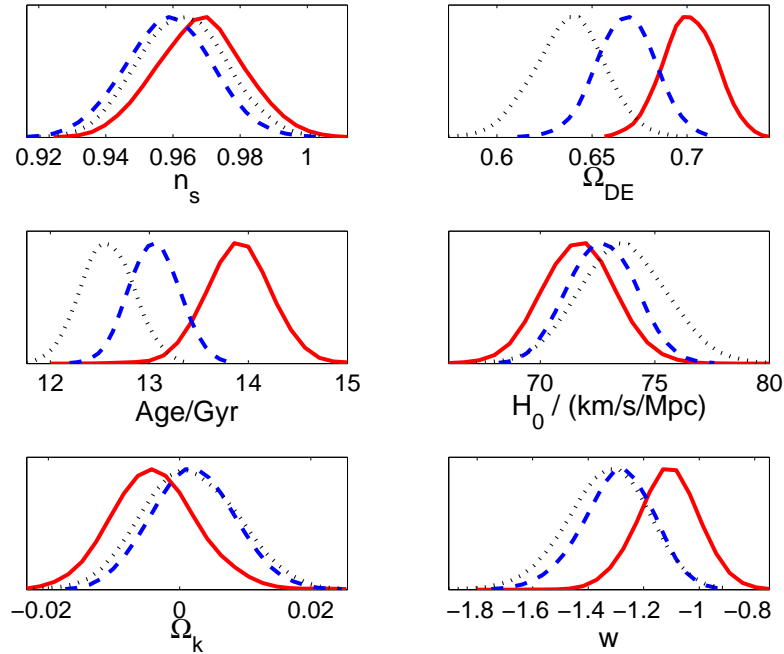}
\caption{Marginalized parameter distributions for wCDM+k with and without additional sterile neutrinos. Labels are the same as in Figure \ref{fig:1D_LCDM}. Since these models are very poorly constrained when using WMAP7 data only, we only show the results from WMAP7++.}
\label{fig:1D_kw}
\end{figure*}

In Figure \ref{fig:1D_kw} and Table \ref{tab:wcdm} we show the corresponding results for the \wcdm{} model, but only for the WMAP7++ data sets. We basically find the same shifts in $\Omega_{\rm DE}$ (corrsponding to $\Omega_\Lambda$) and in the age of the universe, for the same reasons as explained above. When we apply the tight prior on $H_0$ from \citet{riess:2011}, WMAP data will constrain the universe to be close to flat for all models, as can be seen in the resulting limits on $\Omega_k$. 

When including sterile neutrinos, we notice that $w$ is shifted into the $w<-1$ phantom regime.  For decreasing $\Omega_{\rm DE}$,  $w$ is forced to smaller, i.e., more negative values  to obtain the late time acceleration required by the supernova data. 

\begin{figure*}[htb]
\center
\includegraphics[width=1.4\columnwidth]{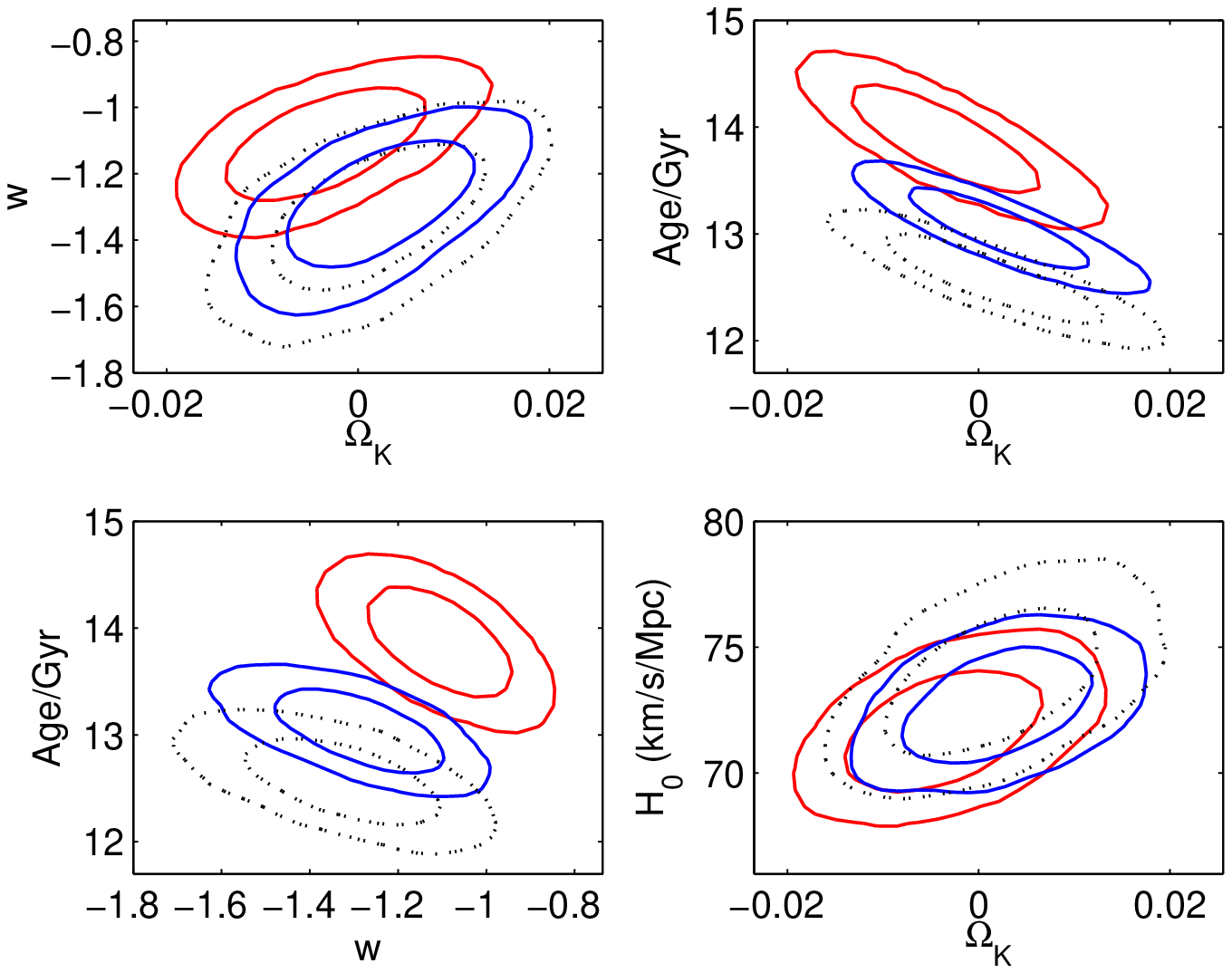}
\caption{Marginalized 2D parameter distributions for wCDM+k with and without additional sterile neutrinos. Labels are the same as in Figure \ref{fig:1D_LCDM}. Since these models are very poorly constrained when using WMAP7 data only, we only show the results from WMAP7++. }
\label{fig:2D_kw}
\end{figure*}

In Figure \ref{fig:2D_kw} we show 2 dimensional 68\% and 95\% confidence contours for a few strongly correlated parameters. We see from the $\Omega_k$-$w$ contours, that for no sterile neutrinos, the \LCDM{} model ($\Omega_k=0$ and $w=-1$)  falls within the 68\% contours, while with sterile neutrinos, the \LCDM{} $+ \nu_{\rm s}$ model falls just outside the 95\% contours. This indicates that, if further oscillation experiments confirm the existence of sterile neutrinos with properties close to what was found in \citet{kopp:2011}, this will imply some tension between the cosmological constant as dark energy and cosmological data. One might argue that models with $w<-1$ are unphysical. However there exist several physically more viable dark energy models that will give an effective equation of state $w<-1$ at late times and thus be able to accommodate the current cosmological data better than the \LCDM{} model in a cosmology with more mass in the neutrino sector (see e.g. \citet{lavacca:2009}). A discussion of different dynamical dark energy models are outside the scope of this paper, but our results indicate  that sterile neutrinos with the properties derived in \citet{kopp:2011} would call the cosmological constant into question as the explanation for dark energy.

%\begin{widetext}
%\end{widetext}

\begin{table*}[htb]
\begin{center}
\begin{tabular}{crlrlrl}
\hline \hline
 Parameter& \multicolumn{2}{c}{\LCDM} &\multicolumn{2}{c}{\LCDM+1$\nu_s$}& \multicolumn{2}{c}{\LCDM+2$\nu_s$} \\
\hline
$n_s$ & \quad0.968 & $\pm$0.014 & \quad 0.959 & $\pm$ 0.013 & \quad0.963 & $\pm$ 0.013 \\
$\Omega_{DE}$ & 0.700 & $\pm$0.014 & 0.668 & $\pm$ 0.015 & 0.638 & $\pm$ 0.018 \\
Age (Gyr) &13.87&$\pm$0.31& 13.05&$\pm$0.25& 12.58&$\pm$0.26 \\
$H_0$ (km/s/Mpc) &71.7&$\pm1.5$&72.7&$\pm$1.4&73.6&$\pm$1.9\\
$\Omega_k$ &-0.0033&$\pm0.0066$&0.0021&$\pm$0.0062&0.0016&$\pm$0.0071\\
$ w$ & { -1.11}&$ {  \pm0.10}$&{ -1.29}& ${ \pm0.12}$&{-1.33}&${ \pm 0.14}$\\
${ \Delta \chi^2}$ & 0 && -12.0  &&  -9.3& \\ 
\hline
\end{tabular}
\caption{One-dimensional marginalized parameter distributions for the \wcdm{} model with and without sterile neutrinos. The errors shown are the 1$\sigma$ deviations from the mean value.  We also show the $\Delta \chi^2$ values for the different models where ${  \Delta \chi^2_{\rm model} = \chi^2_{\Lambda {\rm CDM}} - \chi^2_{\rm model}}$.}
\label{tab:wcdm}
\end{center}
\end{table*}

%\begin{widetext}
%\end{widetext}

{ In Tables \ref{tab:lcdm} and \ref{tab:wcdm} we also show the $\Delta \chi^2$ between the best-fit models in the cases with and without the added massive sterile neutrinos. We see the addition of sterile neutrinos decreases the fit to the data, especially when using the WMAP7++ data sets and the \LCDM{} model, giving $\Delta \chi^2 = -22.6$  for the $2\nu_s$ case. When opening for the $\Omega_k$ and $w$ degrees of freedom, the $\Delta \chi^2$ reduces to -9.3, underlining the need to look beyond a flat \LCDM{} model if the existence of the sterile neutrinos in \citet{kopp:2011} is confirmed. It is interesting to notice, that the worsening of $\chi^2$ from introducing $2\nu_s$ in our analysis, is of the same order as the improvement of the fit by introducing the same sterile neutrinos in \citet{kopp:2011}. As previously stated, our point of departure in this work is that neutrino oscillation experiments are less prone to systematical errors than cosmological observations, justifying the use of the results from \citet{kopp:2011} as an input in the cosmological models.}

\section{Discussion and conclusions}

To summarize, we have investigated how the presence of one or two
sterile neutrinos with the properties estimated in \cite{kopp:2011}
changes the preferred values of cosmological parameters.  Rather than
deriving constraints on neutrino properties from cosmology, we chose
the opposite approach of using neutrino experiments to constrain
cosmology.  We think our approach can be justified, since the
uncertain factors in cosmology, like the nature of the dark energy,
are arguably larger than those in neutrino physics.  

We analyzed the $\Lambda$CDM model and the $w$CDM model 
with spatial curvature as an added parameter, and for two data sets: 
WMAP7 alone, and WMAP7 plus
large-scale structure, supernovae type Ia, and the HST result for the
Hubble parameter.   The most interesting changes from the standard
$\Lambda$CDM model with no sterile neutrinos were in the equation of
state parameter $w$ and in the age of the universe.  

In the \wcdm+$2\nu_s$ model we found a preferred age of $\sim 12.5$Gyr. One might question whether such a young universe would be in conflict with other cosmological observations. Regarding observations of high redshift objects, which ages are derived from their redshifts, this should not be a problem.   As an example, the age of the universe at redshift 12 in the mean parameter value \wcdm+$2\nu_s$ model from WMAP7++ is only 16\% lower than for a standard \LCDM{} model.

One may also put lower limits on the age of the universe by measuring the age of the oldest objects in a cosmology-independent way. A common way to do this is by main sequence fitting in globular clusters. In \citet{Gratton:2003} they use this technique, and estimate the age of the oldest globular cluster in the galaxy to be $13.4 \pm 0.8 \pm 0.6$ Gyr (statistical/systemetic errors), which leaves the 2$\nu_s$ models within the error bars.  { In \citet{Wang:2010} nine cluster with estimated ages of around 14 Gyr were found in M31.}  In \citet{frebel:2007} they use decay rates of radioactive isotopes in nearby stars to estimate a stellar age of 13.4 Gyr. However, the uncertainties in these kinds of measurements are large, and the authors estimate an uncertainty of $\sim$ 2 Gyrs, which also places the model with 2 massive sterile neutrinos well within the allowed range.

We found that the 3 active + 2 sterile neutrinos scenario prefers an
equation of state parameter for dark energy $w < -1$, with the
cosmological constant being ruled out at 2$\sigma$. The fact that $w < -1$ should not be taken as an indication of phantom energy.  It more
likely means that the correct dark energy model cannot be described by
a constant $w$.  If the evidence for sterile neutrinos from
oscillation experiments becomes conclusive the implication could be
that the cosmological constant is ruled out as dark energy.

\begin{acknowledgements}
We thank Frode K. Hansen for useful discussions. The results described in this paper have been produced using the Titan High Performance Computing facilities at the University of Oslo. 
\end{acknowledgements}

\bibliographystyle{aa} % style aa.bst
%\bibliography{cites}

\end{document}